\documentclass[sn-mathphys]{sn-jnl}

\usepackage{braket}
\usepackage{amsfonts}
\usepackage{color}
\usepackage{amssymb}
\usepackage{mathrsfs}
\usepackage{graphicx}
\usepackage{lmodern}
\usepackage{lineno}
\usepackage{hyperref}
\usepackage[normalem]{ulem}
\usepackage{soul}

\newcommand{\beq}{\begin{eqnarray}}
\newcommand{\eeq}{\end{eqnarray}}

\newcommand{\Cyl}{\mathrm{Cyl}}
\newcommand{\nn}{\nonumber}

\jyear{2022}%

\begin{document}



\title[Star product approach for Loop Quantum Cosmology]{Star product approach for Loop Quantum Cosmology}


\author*[1,2]{\fnm{Jasel} \sur{Berra-Montiel}}\email{jasel.berra@uaslp.mx}

\author[1,2]{\fnm{Alberto} \sur{Molgado}}\email{alberto.moglado@uaslp.mx}

\author[1]{\fnm{Eduardo} \sur{Torres-Cordero}}\email{a239602@alumnos.uaslp.mx}

\affil*[1]{\orgdiv{Facultad de Ciencias}, \orgname{Universidad Aut\'onoma de San Luis 
Potos\'{\i}}, \orgaddress{\street{Campus Pedregal, Av. Parque Chapultepec 1610,  Col. Privadas del Pedregal}, \city{San Luis Potos\'{\i}}, \postcode{78217}, \state{SLP}, \country{Mexico}}}

\affil[2]{\orgdiv{Dual CP Institute of High Energy Physics}, \country{Mexico}}

\date{Received: date / Revised version: date}



\abstract{
Guided by recent developments towards the implementation of 
the deformation quantization program within the Loop 
Quantum Cosmology (LQC) formalism, in this paper we address the introduction of both the integral and differential representation of the star product for LQC.  To this end, we consider the  Weyl quantization map for cylindrical functions defined on the Bohr compactification of the reals.   The integral representation 
contains all of the common properties that characterize a star product which, in the case under study here, stands for a deformation of the  usual pointwise product of cylindrical functions. Our construction also admits a direct comparison with the integral representation of the Moyal product which may be reproduced from our formulation by judiciously substituting the appropriate characters that identify such representation.   Further, we introduce a suitable star commutator that correctly reproduces both the quantum representation of the holonomy-flux algebra for LQC and, in the proper limit, the holonomy-flux classical Poisson algebra emerging in the cosmological setup.  Finally, we propose a natural way to obtain the quantum dynamical evolution in LQC in terms of this star commutator for cylindrical functions as well as a differential representation of the star product using discrete finite differences.  We expect that our findings may contribute to a better understanding of certain issues arising within the LQC program, in particular, those related to the semiclassical limit and the dynamical evolution of quantum states.
}

\keywords{Loop Quantum Cosmology, Deformation quantization, star product, Bohr compactification}
\pacs[MSC Classification]{83C45, 81S30, 46F10}


\maketitle

\section{Introduction}

Loop Quantum Cosmology (LQC) is a theory inspired by the quantization techniques developed within the Loop Quantum Gravity (LQG) program to bring together the principles of General Relativity and Quantum Mechanics.  In particular, the LQC framework is associated with finite-dimensional minisuperspace models for which certain symmetries, such as homogeneity and isotropy, for the background spacetimes are assumed~\cite{Bojowald},~\cite{report}. Within this framework, significant advances in the quantum gravity program have been reported, being of particular relevance the analysis of the microscopic ground for black hole entropies \cite{Rovelli}, \cite{Krasnov}, \cite{Domangala}, the avoidance of classical singularities as a consequence of quantum bounces \cite{report}, \cite{bounce}, and the investigation of inhomogeneities imprinted on the cosmic expansion \cite{Bojowald1}, \cite{Agullo}. Despite the remarkable progress on those topics, it is worth mentioning that in recent years considerable work has been dedicated towards the elucidation of certain technical issues, which continue not completely well understood at the present time, in particular, those concerning a detailed evolution of quantum states \cite{CRovelli}, \cite{Perez}, the understanding of an appropriate semiclassical limit \cite{Vandersloot}, \cite{Flori1}, \cite{ELivine}, \cite{Dasgupta}, \cite{Liu}, the emergence of effective theories \cite{MBojowald1}, \cite{MBojowald2}, \cite{MBojowald3} and the discussion of quantization ambiguities \cite{APerez}, \cite{Brahma}, \cite{Navascues}, among others.  These topics remain major open questions and any contribution towards their understanding may undoubtedly add
to validate a realistic description of the LQC theory. (We refer the reader to~\cite{Critics} for a detailed review on these subjects from a modern perspective.)

In order to shed some light on the issues mentioned above, a different approach for the LQG program has been recently proposed \cite{Sahlmann}, \cite{Perlov}, \cite{DQPolymer}, \cite{PolyW}, \cite{Quasi}, \cite{Tomography}, (see also \cite{Stottmeister1}, \cite{Stottmeister2}, \cite{Stottmeister3} for a Born-Oppenheimer approach to the space adiabatic perturbation theory). This novel approach is based on the deformation quantization picture of quantum mechanics. Deformation quantization, also referred to as phase space quantum mechanics in the literature, consists of an alternative quantization procedure based on the idea that a quantum system is obtained by deforming the algebraic and
geometrical structures defined in the classical phase space \cite{Bayen1}, \cite{Bayen2}. Undeniable, one of the most prominent features of the deformation quantization approach is related to the algebra of quantum observables, in consideration of the fact that this algebra is not specified by a set of self-adjoint operators acting on a Hilbert space, as in ordinary quantum mechanics, but instead the observables are characterized by complex-valued functions defined on the classical phase space for which the conventional commutative pointwise product is replaced by an associative, non-commutative product, the so-called star product. As a consequence, the introduction of the star product induces a deformation of the Poisson algebra in such a way that the information contained in the quantum commutators of any pair of self-adjoint operators is mapped to the deformed algebraic classical structures. Until now, the formalism of deformation quantization has not only provided significant contributions in pure and applied mathematics \cite{Kontsevich}, \cite{Waldman}, but it also has proved to be an outstanding tool for the quantum analysis of a broad variety of physical systems \cite{Fredenhagen}, \cite{Compean}, \cite{Compean2}, recently including the treatment of constrained systems \cite{GA}, the coherent field quantization \cite{coherent}, \cite{path}, and the tomographic representation for fields \cite{tomographic}.  

Hence, the main aim of this  manuscript is to extend previous results obtained within the deformation quantization framework to the LQG context by explicitly addressing both the integral and differential representations of the star product for LQC. To this end, 
we start by considering the Weyl quantization map on a phase space given by the cylindrical functions defined on the Bohr compactification of the reals and, in complete analogy 
to the Moyal product, by conveniently introducing an associative and non-commutative star product for LQC that accounts for an appropriate deformation of the pointwise product.  As a consequence, we are able to identify an integral representation for this star product, which in turn allows us to define a suitable star commutator  that not only reproduces the correct quantum representation of the holonomy-flux algebra for LQC, but it also duplicates, in the relevant limit, the Poisson algebra emerging within the classical cosmology scenario. Then, by using a discrete version of the Taylor expansion, derived from the calculus of finite differences, a candidate of a differential representation of the star product is obtained. Our claim is thus, that such representations for the star product in LQC may allow us to analyze, in a more succinct manner, several aspects related to both the semiclassical limit and the analysis of quantization ambiguities commonly encountered within the LQC framework. We also hope that the proposed star product for LQC, together with the introduction of a family of quasi-probability distributions, as developed in~\cite{Quasi},  would be advantageous in order to address, from an innovative perspective, the problem associated with the dynamical evolution of quantum states in LQC. Notably, determining these quasi-probability distributions may enable us to analyze several aspects associated with squeezed states, which in turn may provide an appropiate non-perturbative vacuum in LQC, as stated in \cite{squeezed}, \cite{Corichi}.  Further, the phase space description may allow us to investigate the correspondence between classical and quantum coherence theory, thus playing a fundamental issue within the LQC framework for testing the semi-classical limit \cite{Oriti}. Finally, the definition of a star product allow us to obtain both the physical inner product and the physical Wigner distribution, the latter corresponding to the phase space representation of the density operator \cite{GA}. This quasi-probability distribution encodes all of the dynamical information by means of star-genvalue equations, and it is also responsible for all the auto-correlation properties and the transition amplitudes of a given quantum mechanical system, since even for chaotic systems, the quantum evolution can be depicted by following contour plots of different phase space distributions~\cite{Cosmas}, \cite{Lee}, \cite{Gradoni}.

The paper is organized as follows: in section 2 we briefly review the basic concepts behind the Weyl quantization scheme for functions defined on the Bohr compactification of the reals. In section 3, we introduce the integral representation of the star product for LQC and derive some of its properties. In particular, in analogy with the Moyal's prescription for quantum mechanics,
we define the star commutator and, by means of the Wigner distribution for LQC, 
we establish the dynamical evolution for quantum systems, the uncertainty principle and the Ehrenfest's theorem. In section 4, using some tools from calculus of finite differences, a differential representation of the star product in LQC is presented. Finally, we conclude in section 5 with some comments and perspectives about our findings. 

\section{The Wigner-Weyl correspondence in LQC}
In order to obtain the integral and differential representations of the star product for LQC, in this section, we briefly review the definition of the Weyl quantization map for the Bohr compactification of the reals, $b\mathbb{R}$. First, we start by considering some basic features of harmonic analysis regarding the analytical properties of functions defined on $b\mathbb{R}$ (for further details on this subject we refer the reader to \cite{Folland}, \cite{Reiter}, \cite{Shubin}). For the sake of simplicity, we restrict our attention to systems with one degree of freedom, although its generalization to a higher dimensional setup follows straightforwardly.

Let $\mathbb{R}$ be a locally compact Abelian group, given by the real numbers equipped with the usual  addition operation and the standard topology. This means, that the real numbers $\mathbb{R}$ forms a metrizable, $\sigma$-compact and locally compact Abelian group.  In particular, the $\sigma$-compact property implies that there exists a sequence $K_{n}\subset K_{n+1}$ of compact subsets, such that $\mathbb{R}=\cup_{n}K_{n}$, i.e., the real numbers $\mathbb{R}$ comprises a compact exhaustion \cite{Deitmar}. A character  of the reals $\mathbb{R}$, is given by a continuous group homomorphism $h_{\mu}: \mathbb{R}\to\mathbb{T}$, to  the unit torus, $\mathbb{T}=\{z\in\mathbb{C}:|z|=1\}$, so $h_{\mu}$ is a map satisfying
\begin{equation}
h_{\mu}(a+b)=h_{\mu}(a)h_{\mu}(b),
\end{equation}
for every $a,b\in\mathbb{R}$. Let us denote by $\widehat{\mathbb{R}}$, the set of all characters of $\mathbb{R}$, labeled by $\mu\in\mathbb{R}$,
\begin{equation}\label{character}
h_{\mu}(a)=e^{i\mu a}.
\end{equation}
By using the $\sigma$-compact property of the reals, one can define a uniform metric on $\widehat{\mathbb{R}}$ such that it endows the set of characters with the properties of a locally compact group, called the dual group of $\mathbb{R}$ \cite{Deitmar}. The map given by (\ref{character}) defines a group isomorphism $h_{\mu}:\mathbb{R}\to\widehat{\mathbb{R}}$, hence, we can say that $\widehat{\mathbb{R}}$ is isomorphic to $\mathbb{\mathbb{R}}$ as locally compact groups. Let us consider the dual group $\widehat{\mathbb{R}}$, but now equipped with the discrete topology $\widehat{\mathbb{R}}_{discr}$. By definition, this group turns out to be discrete and Abelian. Then, by making use of the Pontryagin's duality theorem, its dual group, denoted by $\widehat{\widehat{\mathbb{R}}}_{discr}$, forms a compact Abelian group, commonly named as the Bohr compactification of the reals, $b\mathbb{R}$. On $b\mathbb{R}$, the group of the real numbers $\mathbb{R}$ can be embedded as a dense subgroup and the addition operation is extended uniquely to the continuous group operation given by the characters \cite{Folland}. Finally, by applying the Pontryagin's duality theorem once more, the dual group associated with the Bohr compactification, $\widehat{b\mathbb{R}}$, proves to be discrete, which means that no continuity requirements are demanded on its characters since this group is equipped with the discrete topology.

Seen as locally compact groups, both the Bohr compactification, $b\mathbb{R}$, and 
its dual group, $\widehat{b\mathbb{R}}$, carry a unique normalized Haar measure,
respectively.  Thus,  the Haar measure $dc$ on $b\mathbb{R}$ is described by
\begin{equation}\label{HaarbR}
\int_{b\mathbb{R}}dc\,h_{\mu}(c)=\delta_{\mu,0} \,,
\end{equation}
for any character $h_{\mu}$, while the Haar measure $d\mu$ defined on $\widehat{b\mathbb{R}}$  
explicitly reads
\begin{equation}\label{HaardualbR}
\int_{\widehat{b\mathbb{R}}}d\mu\tilde{f}_{\mu}=\sum_{\mu\in\mathbb{R}}\tilde{f}_{\mu} \,,
\end{equation} 
and, hence, it corresponds to the counting measure on $\mathbb{R}$.  
Here $\tilde{f}_{\mu}$ stands for the Fourier transform  on $b\mathbb{R}$, 
that allows us to determine an isomorphism between $L^{2}(b\mathbb{R},dc)$ and $L^{2}(\widehat{b\mathbb{R}},d\mu)$  such that
\begin{equation}\label{FourierbR} 
\tilde{f}_{\mu}=\int_{b\mathbb{R}}dc\,f(c)h_{-\mu}(c).
\end{equation}
By the Peter-Weyl theorem \cite{Folland}, the characters $h_{\mu}$ form an orthonormal uncountable basis for $L^{2}(b\mathbb{R},dc)$, resulting in a non-separable Hilbert space structure. Furthermore, the Hilbert space $L^{2}(b\mathbb{R},dc)$ is isomorphic to the Hilbert, $B^2(\mathbb{R})$, given by the Besicovitch almost periodic functions on $\mathbb{R}$~\cite{Chojnacki}.

In order to define the Weyl quantization map on the Bohr compactification of the reals, let us define the set of cylindrical functions, denoted by $\Cyl(b\mathbb{R})$, as the finite span of characters on $b\mathbb{R}$, and by $\Cyl(\widehat{b\mathbb{R}})$ the image of $\Cyl(b\mathbb{R})$ under the Fourier transform. This means that any $\psi\in \Cyl(b\mathbb{R})$ can be expressed in the form
\begin{equation}\label{sum}
\psi(c)=\sum_{\mu}\tilde{\psi}_{\mu}h_{\mu}(c),
\end{equation}
where $\tilde{\psi}_{\mu}$ represents the Fourier coefficient 
\begin{equation}
\tilde{\psi}_{\mu}=\int_{b\mathbb{R}}dc\,\psi(c)\overline{h_{\mu}(c)},
\end{equation}
which vanishes for all but countable $\mu\in\mathbb{R}$. Therefore, the discrete sum depicted in (\ref{sum}) indicates that the space $\Cyl(b\mathbb{R})$ is given by the set of all complex valued functions on $\mathbb{R}$ that are vanishing almost everywhere except for a countable number of points.

It is worth mentioning here that the central difference between the Schr\"odinger representation of quantum mechanics and LQC lies in the choice of the Hilbert space. While, in the Schr\"odinger representation the Hilbert space is denoted by $L^{2}(\mathbb{R})$, in LQC the Hilbert space is commonly designated as $L^{2}(b\mathbb{R},dc)$. However,  a distinctive difference between these Hilbert spaces relies on the fact that the position and momentum operators $\hat{q}$ and $\hat{p}$ are well-defined operators on $L^{2}(\mathbb{R})$, whereas in LQC this is no longer true.  The reason stems from the full LQG theory, in which case a Hermitian operator corresponding to the analogue of the position operator, namely the gauge connection, does not exist~\cite{Ashtekar},~\cite{Singh}. Nevertheless, since the holonomies associated with those connection variables are well-defined, instead one 
may consider as fundamental variables the operators $\hat{h}_{\mu}$ and $\hat{p}$ given by
\begin{equation}\label{hp}
\hat{h}_{\mu}\psi(c):=h_{\mu}(c)\psi(c) \,, \hspace{5ex} \hat{p}\psi(c):=\sum_{\mu\in\mathbb{R}}\hbar\mu\tilde{\psi}_{\mu}h_{\mu}(c) \,,
\end{equation}
where $\psi\in \Cyl(b\mathbb{R})$. In particular, one may observe from the first of these definitions and by using property (\ref{HaarbR}), that $h_{\mu}(c)\psi(c)$ proves to be non-differentiable on $L^{2}(b\mathbb{R},dc)$. This prevents the existence of a position operator on the Hilbert space $L^{2}(b\mathbb{R},dc)$, thus rendering a different representation of Quantum Mechanics by evoking the Stone-von Neumann uniqueness theorem~\cite{Takhtajan}.
In consequence, the operators defined in~(\ref{hp}) provide the basic commutator relations
\begin{eqnarray}\label{hfalgebra}
\left[\hat{h}_{\mu},\hat{p} \right]\psi(c)&=&-\hbar\mu h_{\mu}\psi(c), \nonumber \\
\left[\hat{h}_{\mu},\hat{h}_{\nu} \right]\psi(c)&=&0=\left[\hat{p},\hat{p} \right]\psi(c),  
\end{eqnarray}
for all $\psi\in L^{2}(b\mathbb{R},dc)$, which correspond to the LQC analogue of the quantum representation of the holonomy-flux algebra in LQG \cite{report}.

Given a distribution $g$ on $\Cyl(b\mathbb{R}\times \widehat{b\mathbb{R}})$ (also called a symbol, according to the terminology adopted in harmonic analysis \cite{Folland}), that is, $g\in \Cyl(b\mathbb{R}\times \widehat{b\mathbb{R}})^{*}$, where $\Cyl(b\mathbb{R}\times \widehat{b\mathbb{R}})^{*}$ denotes the dual space of $\Cyl(b\mathbb{R}\times \widehat{b\mathbb{R}})$ \cite{Sahlmann}, and making use of the momentum representation with the purpose to avoid the issue of multiplying an element of $b\mathbb{R}$ by an arbitrary real number, the quantization map for LQC of the function $g(c,\mu)$, in the Weyl symmetric ordering, is defined as \cite{Sahlmann}, \cite{Quasi},
\begin{equation}\label{WeylLQC}
\hat{g}\psi(c)=\mathcal{Q}^{LQC}(g)\psi(c):=\int_{\widehat{b\mathbb{R}}\times \widehat{b\mathbb{R}}}d\mu d\nu\,\tilde{g}\left(\mu-\nu, \hbar\left( \frac{\mu+\nu}{2}\right) \right)h_{\mu}(c)\tilde{\psi}_{\nu} \,, 
\end{equation} 
where $\psi\in L^{2}(b\mathbb{R},dc)$, and $\tilde{g}$ denotes the partial Fourier transform of the function $g(c,\mu)$ with respect to the first variable. Following~\cite{Quasi}, after some straightforward calculations, we may demonstrate that the quantization map $\mathcal{Q}^{LQC}$, defined in (\ref{WeylLQC}), uniquely determines the holonomy and momentum operators stated in (\ref{hp}), that is, 
\begin{equation}
\mathcal{Q}^{LQC}(h_{\mu}(c))\psi(c)=\hat{h}_{\mu}\psi(c), \;\;\;\; \mathcal{Q}^{LQC}(\mu)\psi(c)=\hat{p}\psi(c), 
\end{equation} 
and 
\begin{equation}
\mathcal{Q}^{LQC}(\mu h_{\mu}(c))\psi(c)=\frac{1}{2}\left(\hat{h}_{\mu}\hat{p}+\hat{p}\hat{h}_{\mu} \right)\psi(c), 
\end{equation}
which, as we can observe, corresponds to the Weyl-symmetric ordering quantization prescription for LQC \cite{Sahlmann}.

\section{Star product representation of LQC}
Let $f$ and $g$ denote functions acting on $\Cyl(b\mathbb{R}\times \widehat{b\mathbb{R}})$, then, by means of the  Weyl quantization map for LQC defined in the previous section, $\mathcal{Q}^{LQC}(f)$ and $\mathcal{Q}^{LQC}(g)$ correspond to operators acting on the Hilbert space $L^{2}(b\mathbb{R},dc)$. Since the product $\mathcal{Q}^{LQC}(f)\mathcal{Q}^{LQC}(g)$ also determines an operator on $L^{2}(b\mathbb{R},dc)$, in this section we introduce the so-called star product of symbols defined by
\begin{equation}
\mathcal{Q}^{LQC}(f)\mathcal{Q}^{LQC}(g)=\mathcal{Q}^{LQC}(f\star_{LQC} g),
\end{equation}
where the star product $\star_{LQC}$, like the quantization map $\mathcal{Q}^{LQC}$ itself, explicitly depends on $\hbar$. In order to compute $f\star_{LQC} g$, let us first express the Weyl quantization map (\ref{WeylLQC}) in terms of an integral operator as
\begin{equation}
\hat{g}\psi(c)=\int_{b\mathbb{R}}db\,K_{g}(c,b)\psi(b),
\end{equation}
where the integral kernel $K_{g}(c,b)$ is given by
\begin{equation}\label{kernel}
K_{g}(c,b)=\int_{\widehat{b\mathbb{R}}\times \widehat{b\mathbb{R}}}d\mu d\nu\, \tilde{g}(\mu,\nu)h_{\frac{\mu}{2}}(c+b)h_{\frac{\nu}{\hbar}}(c-b).
\end{equation}
This integral is not necessarily absolutely convergent, and should be understood as a partial Fourier transform with respect to $L^{2}(\widehat{b\mathbb{R}},d\mu)$. We can use (\ref{kernel}) to recover the partial Fourier transform of the symbol $g$ from the kernel $K_{g}(c,b)$ by noting that
\begin{equation}
    \tilde{K}_g (\alpha,\beta)=\tilde{g}\left(\alpha+\beta,\hbar\left( \frac{\alpha-\beta}{2}\right) \right),
\end{equation}
where $\tilde{K}_{g}(\alpha,\beta)$ denotes the Fourier transform of the integral kernel, and hence it is easy to see that
\begin{equation}\label{pFourier}
\tilde{g}(\mu,\nu)=\int_{b\mathbb{R}^{2}\times \widehat{b\mathbb{R}}^2}da\,db\,d\alpha\,d\beta\, \tilde{K}_{g}(\alpha,\beta)h_{\alpha+\beta-\mu}(a)h_{\frac{\hbar}{2}\left( \alpha-\beta\right) -\nu}(b).
\end{equation}
Now, we can verify, using the Weyl quantization map stated in (\ref{WeylLQC}), that the action given by the product of the two operators $\hat{f}$, $\hat{g}$ on any $\psi\in L^{2}(b\mathbb{R},dc)$ can be written as
\begin{eqnarray}
\label{ProductOp-kernel}
(\hat{f}\hat{g})\psi(c)&=& \mathcal{Q}^{LQC}(f)\mathcal{Q}^{LQC}(g)\psi(c) \nonumber \\
&=& \int_{b\mathbb{R}^{2}\times \widehat{b\mathbb{R}}^4}  da\,db\,d\alpha\,d\beta\,d\mu\,d\nu\tilde{f}(\alpha,\beta)\tilde{g}(\mu,\nu)\nonumber h_{\frac{\alpha}{2}}(c+b)\nonumber\\ && \times h_{\frac{\mu}{2}}(a+b)h_{\frac{\beta}{\hbar}}(c-b)h_{\frac{\nu}{\hbar}}(b-a)\psi(a) \nonumber \\
&=&\int_{b\mathbb{R}}da\,K_{fg}(c,a)\psi(a),
\end{eqnarray}
where in this case, the integral kernel reads
\beq
\mkern-100mu K_{fg}(c,a)
& = & \int_{b\mathbb{R}\times \widehat{b\mathbb{R}}^4}db\,d\alpha\,d\beta\,d\mu\,d\nu\, \left[\tilde{f}(\alpha,\beta)\tilde{g}(\mu,\nu)\right.
\nn\\
& & 
\times \left. h_{\frac{\alpha}{2}}(c+b)h_{\frac{\mu}{2}}(a+b)h_{\frac{\beta}{\hbar}}(c-b)h_{\frac{\nu}{\hbar}}(b-a) \right] \,.
\label{ProductKernel}
\eeq
Substituting this expression on (\ref{pFourier}), and after computing several integrations with properties (\ref{HaarbR}) and (\ref{HaardualbR}), we obtain 
\begin{equation}\label{Fstar}
(\widetilde{f\star_{LQC}}g)(c,\mu)=\int_{b\mathbb{R}\times \widehat{b\mathbb{R}}}da\,d\mu\,\tilde{f}(c-a,\mu-\nu)\tilde{g}(a,\nu)h_{-\frac{\hbar\nu}{2}}(c)h_{\frac{\hbar\mu}{2}}(a).
\end{equation} 
and by using property (\ref{FourierbR}), the symbol associated to the product of the operators $\hat{f}$ and $\hat{g}$, denoted as $f\star_{LQC} g$, is obtained as follows
\beq
\mkern-90mu(f\star_{LQC} g)(c,\mu) 
& = & 
\int_{b\mathbb{R}^{2}\times \widehat{b\mathbb{R}}^{2}}da\,db\,d\rho\,d\nu \, \left[f(a,\rho)g(b,\nu) \right.
\nn\\
&   & 
\times \left. h_{\frac{2}{\hbar}(\rho-\nu)}(c)h_{\frac{2}{\hbar}(\mu-\rho)}(b)h_{\frac{2}{\hbar}(\nu-\mu)}(a)\right]
\,.
\label{star}
\eeq
This non-commutative bilinear product between two symbols $f(c,\mu),\ g(c,\mu) \in \Cyl(b\mathbb{R}\times \widehat{b\mathbb{R}})^{*}$, corresponds to the integral representation of the star product in the context of LQC, and coincides with the operator product obtained in \cite{Stottmeister2}, where a coherent Segal-Bargmann-Hall transform for compact Lie groups was employed. However, we must emphasize here that, within our formulation, the integral representation of the star product was obtained straightforwardly, and in complete correspondence with the standard integral representation for the Moyal product in Quantum Mechanics in phase space, thus avoiding any additional mathematical structure other than the inherent to the Bohr compactification and its respective characters.

Moreover, as a consequence of the associativity of the product of operators on a Hilbert space, that is, $(\hat{f}\hat{g})\hat{h}=\hat{f}(\hat{g}\hat{h})$, one may easily demonstrate that the star product (\ref{star}) also defines an associative product for the symbols, that is,
\begin{equation}
\left( (f\star_{LQC} g)\star_{LQC} h\right)  (c,\mu)=\left( f \star_{LQC}(g\star_{LQC} h)\right) (c,\mu).
\end{equation}
Besides, it is worth noticing that if in the integral formula (\ref{star}) we replace the characters $h_{\mu}(c)$ of the Bohr compactification of the reals $b\mathbb{R}$ with the characters associated with the group of the real numbers $\mathbb{R}$, which in the latter case are merely given by exponentials,  we may recognize that the star product stated in  (\ref{star}) corresponds to the integral representation of the Moyal product obtained in ordinary quantum mechanics \cite{Zachos}. This means that the star product $\star_{LQC}$ may be thought of as a generalization of the Moyal product between functions defined on the Schwartz space $\mathcal{S}(\mathbb{R})$ \cite{Takhtajan} by allowing the incorporation of symbols defined on $\Cyl(b\mathbb{R}\times \widehat{b\mathbb{R}})^{*}$.   

In complete analogy to the Moyal product, the star product $\star_{LQC}$ thus determines a deformation of the ordinary pointwise product of functions on $\Cyl(b\mathbb{R}\times \widehat{b\mathbb{R}})^{*}$. In order to appreciate such deformation, note that if we set $\hbar\to 0$, the formula for $(\widetilde{f\star_{LQC}}g)(c,\mu)$, depicted in (\ref{Fstar}), reduces to the convolution of $\tilde{f}$ and $\tilde{g}$, which is equal to the Fourier transform of the pointwise product $fg$. This means that
\begin{equation}\label{hzero}
\lim_{\hbar\to 0}(f\star_{LQC}g)(c,\mu)=(fg)(c,\mu) \,,
\end{equation}
and thus our claim is valid.

Next, for an arbitrary couple of functions on $f, g\in\Cyl(b\mathbb{R}\times \widehat{b\mathbb{R}})^{*}$ we may define the star commutator as $[f,g]_{\star_{LQC}}:=f\star_{LQC}g-g\star_{LQC}f$.  In particular, 
by choosing the functions $f(c,\mu)=h_{\mu}(c)$ and $g(c,\mu)=\mu$,  we can observe that
\begin{eqnarray}\label{commutator}
\left[ h_{\mu}(c),\mu\right]_{\star_{LQC}}=-\hbar \mu h_{\mu}(c), \nonumber \\
\left[ h_{\mu}(c),h_{\nu}(c)\right]_{\star_{LQC}}= 0=\left[ \mu,\nu\right]_{\star_{LQC}},
\end{eqnarray}
which is nothing but the quantum representation of the holonomy-flux algebra depicted in (\ref{hfalgebra}). Furthermore, by means of the property (\ref{hzero}) of the star product $\star_{LQC}$, the star commutator satisfies
\beq
\label{Poisson}
\lim_{\hbar\to 0}\frac{1}{i\hbar}\left[ h_{\mu}(c),\mu\right]_{\star_{LQC}} & = & 
i\mu h_{\mu}(c)=\left\lbrace h_{\mu}(c),\mu\right\rbrace \,, \nonumber \\
\lim_{\hbar\to 0}\frac{1}{i\hbar}\left[ h_{\mu}(c),h_{\nu}(c)\right]_{\star_{LQC}} & = &  0=\left\lbrace h_{\mu}(c),h_{\nu}(c)\right\rbrace\,, \nn\\ 
 \lim_{\hbar\to 0} \frac{1}{i\hbar}\left[ \mu,\nu\right]_{\star_{LQC}} & = & 
 0=\left\lbrace \mu,\nu\right\rbrace\,. 
\end{eqnarray}
It should be noted that the right hand side of the identities (\ref{Poisson}) agrees with the holonomy-flux Poisson algebra occurring in classical cosmology \cite{report}. Thus, we conclude that the Weyl quantization map $ \mathcal{Q}^{LQC}$ defined in (\ref{WeylLQC})  and the star product $\star_{LQC}$, provides a one-parameter, associative deformation of the holonomy-flux classical algebra. Within the deformation quantization approach \cite{Zachos}, it is expected that the induced deformations become irrelevant at large scales, hence providing a natural mechanism to understand
how the classical description of mechanics emerges from the quantum world.

Finally, in order to establish a  complete parallelism with Moyal's prescription 
for quantum mechanics, we propose that the quantum dynamical evolution in LQC is 
governed by the natural extension of Moyal's equation~\cite{Zachos}
that incorporates the $\star_{LQC}$ bracket defined above and thus, within this context, we introduce the LQC equation  
\begin{equation}\label{dynamical}
\frac{\partial \rho^{LQC}(c,\mu)}{\partial t}=\frac{1}{i\hbar}\left[H(c,\mu),\rho(c,\mu) \right]_{\star_{LQC}}, 
\end{equation} 
where $\rho^{LQC}(c,\mu)$ corresponds to the Wigner quasi-probability distribution for LQC~\cite{Sahlmann}, \cite{Quasi},  while $H(c,\mu)\in\Cyl(b\mathbb{R}\times\widehat{b\mathbb{R}})^{*}$ stands for the classical Hamiltonian. For a given state $\psi\in L^{2}(b\mathbb{R},dc)$, the Wigner distribution $\rho^{LQC}(c,\mu)$ consists of the complex valued function on $b\mathbb{R}\times \widehat{b\mathbb{R}}$, given by
\begin{equation} \label{Wigner}
\rho^{LQC}(c,\mu)=\int_{\widehat{b\mathbb{R}}}d\nu\,\overline{\widetilde{\psi}}_{\mu-\nu/2}\widetilde{\psi}_{\mu+\nu/2}h_{\nu}(c). 
\end{equation}
Note that the dynamical equation for LQC (\ref{dynamical}), as it happens for the Moyal equation, is an extension of Liouville's theorem in classical mechanics and results eminently comparable to the Heisenberg's equation of motion in quantum mechanics \cite{Zachos}, with the significant exception that for the case of our interest here $H(c,\mu)$ and $\rho^{LQC}(c,\mu)$ are functions taking values on $\Cyl(b\mathbb{R}\times \widehat{b\mathbb{R}})$, instead of operators in a Hilbert space.  Further, the star product $\star_{LQC}$  in the LQC equation embodies all of the non-commutative behavior that characterizes the quantum system.  Thus, we claim that within the LQC framework, equation~(\ref{dynamical}) is utterly fundamental in order to elucidate the concealed link between quantum operators and the Poisson structure. Further research is needed to address this last issue.

Finally, both the obtained LQC star product and the Wigner distribution within the LQC framework comprise analogous properties to the usual Moyal product and Wigner function in Quantum Mechanics in phase space~\cite{Zachos}. One may easily check that the star product (\ref{star}) reduces to the pointwise product if integrated over all phase space,
\begin{equation} \label{phasespacestar}
    \int_{b\mathbb{R}\times\widehat{b\mathbb{R}}}  dc \, d\mu \, f \star_{LQC} g = \int_{b\mathbb{R}\times\widehat{b\mathbb{R}}}  dc \, d\mu \, fg = \int_{b\mathbb{R}\times\widehat{b\mathbb{R}}}  dc \, d\mu \, g \star_{LQC} f,
\end{equation}
for any $f$ and $g$ on $\Cyl(b\mathbb{R}\times\widehat{b\mathbb{R}})^*$, whilst, the complex conjugation of the star product reads
\begin{equation} \label{complexstar}
    \overline{(f \star_{LQC} g)}(c,\mu) = (\overline{g} \star_{LQC} \overline{f} ) (c,\mu).
\end{equation}
Additionally, the Wigner distribution as defined in (\ref{Wigner}), turns out to be manifestly real \cite{Quasi} and also satisfies the identity
\begin{equation} \label{Wignerstar}
    (\rho^{LQC} \star_{LQC} \rho^{LQC})(c,\mu) = \rho^{LQC} (c,\mu).
\end{equation}
These properties will be essential to deduce, within the context of deformation quantization, the corresponding uncertainty principle and the Ehrenfest's theorem for LQC in the next section.

\section{The uncertainty principle and Ehrenfest's theorem}
\label{sec:uncertainty}

Now, we discuss the uncertainty principle and the analogue of Ehrenfest's theorem within the context of LQC. For any distribution $f$ on $\Cyl(b\mathbb{R}\times\widehat{b\mathbb{R}})$, the expectation value of the corresponding operator $\hat{f} = \mathcal{Q}^{LQC}(f)$ acting on $L^{2}(b\mathbb{R},dc)$, can be expressed as \cite{Quasi}
\begin{equation} \label{expectation}
    \langle f \rangle = \int_{b\mathbb{R}\times\widehat{b\mathbb{R}}}  dc \, d\mu \, f(c,\mu)\rho^{LQC}(c,\mu) = \left\langle \psi, \hat{f}\psi \right\rangle.
\end{equation}
From the integral representation of the star product depicted in (\ref{star}), using properties (\ref{phasespacestar}), (\ref{complexstar}) and (\ref{Wignerstar}), together with the associative nature of the star product, it follows that the expectation value satisfies
\begin{equation}
    \langle \overline{f} \star_{LQC} f \rangle \geq 0,
\end{equation}
which, in the operator Hilbert space formalism, corresponds to the positiveness property of the norm \cite{Zachos}. In order to obtain the Heisenberg's uncertainty relation, it is convenient to choose \cite{Cosmas} 
\begin{equation}
    f(c,\mu) = a + b \left( \frac{1}{2i\lambda}(h_{\lambda}(c) - h_{-\lambda}(c))\right) + c \mu, 
\end{equation}
where $a,b,c\in \mathbb{C}$ and the second term constitutes the regularized position defined in terms of the characters
\begin{equation}
    q_{\lambda} =\frac{1}{2i\lambda}(h_{\lambda}(c) - h_{-\lambda}(c)) = \frac{\sin(\lambda c)}{\lambda},
\end{equation}
where we have considered $\hbar=1$ for simplicity. The reason for this definition is a consequence of the LQC representation, since the Hilbert space is given by $L^{2}(b\mathbb{R},dc)$, the position operator does not exist, then it has to be written in terms of the characters $h_\mu(c)$, as discussed above (see also \cite{DQPolymer} for further details). The expectation value results in a positive quadratic form
\beq
   0\leq \langle \overline{f} \star_{LQC} f \rangle &=& \overline{a}a + \overline{b}b \langle q_\lambda \star_{LQC} q_\lambda \rangle + \overline{c}c \langle \mu \star_{LQC} \mu \rangle + (\overline{a}b + \overline{b}a) \langle q_\lambda \rangle \nonumber \\
    & & + (\overline{a}c + \overline{c}a) \langle \mu \rangle + \overline{c}b \langle \mu \star_{LQC} q_\lambda \rangle + \overline{b}c \langle q_\lambda \star_{LQC} \mu \rangle \,.
\end{eqnarray}
Now, by using the standard deviation of a quantum operator $(\Delta f)^2 = \langle ( f - \langle f \rangle )^2 \rangle$, and the following relations, derived from the noncommutative star product obtained in (\ref{star}),
\beq
    q_\lambda \star_{LQC} q_\lambda = q_\lambda^2\,, & \qquad & \mu \star_{LQC} \mu = \mu^2 \,,\nonumber \\
    q_\lambda \star_{LQC} \mu = q_\lambda \mu - \frac{1}{2i} \cos(\lambda c)\,, & \qquad &  \mu \star_{LQC} q_\lambda = q_\lambda \mu +\frac{1}{2i} \cos(\lambda c) \,,
\eeq
we have
\begin{eqnarray}
    (\Delta q_\lambda )^2 (\Delta \mu )^2 &\geq& \frac{1}{4} \langle \cos(\lambda c) \rangle^2 + \langle ( q_\lambda - \langle q_\lambda \rangle )( \mu - \langle \mu \rangle ) \rangle^2 \nonumber \\
    &\geq& \frac{1}{4} \langle \cos(\lambda c) \rangle^2,
\end{eqnarray}
and hence
\beq
    \Delta q_\lambda \Delta \mu \geq \frac{1}{2} \mid \langle \cos(\lambda c ) \rangle \mid \,.
\eeq
From this last expression it is straightforward to verify that if we take a small parameter $\lambda$, we obtain
\begin{equation}
    \Delta q_\lambda \Delta \mu \geq \frac{1}{2} \left( 1 - \frac{1}{2} \lambda^2 \Delta q_\lambda^2 + \mathcal{O}(\lambda^4) \right),
\end{equation}
where we have considered the condition $\langle q_\lambda \rangle = 0$. This result agrees with the uncertainty relation stemming from the context of polymer quantum mechanics and other generalized uncertainty principle (GUP) scenarios \cite{DQPolymer}, \cite{Hossain}. 

Finally, let us assume that both $f(c,\mu)$ and the Wigner distribution,
$\rho^{LQC}(c,\mu)$, are explicitly time dependent, then, as claimed before, the dynamical evolution of this observable is specified by the LQC analogue of the Moyal's equation (\ref{dynamical}). From the definition of the expectation value in terms of the Wigner distribution $\rho^{LQC}(c,\mu)$ presented in (\ref{expectation}), we notice
\begin{eqnarray}
    \frac{d}{dt} \langle f \rangle &=& \int_{b\mathbb{R}\times\widehat{b\mathbb{R}}} dc \, d\mu \, \left( \frac{\partial f}{\partial t} \rho^{^{LQC}} + f \frac{\partial \rho^{^{LQC}}}{\partial t} \right) \nonumber \\
    &=& \left\langle \frac{\partial f}{\partial t} \right\rangle + 
    \frac{1}{i\hbar} \int_{b\mathbb{R}\times\widehat{b\mathbb{R}}} dc \, d\mu \, f\star_{LQC}[H,\rho^{^{LQC}}]_{\star_{LQC}}   \nonumber \\
    &=& \left\langle \frac{\partial f}{\partial t} \right\rangle + \frac{1}{i\hbar} \int_{b\mathbb{R}\times\widehat{b\mathbb{R}}} dc \, d\mu \, \rho^{^{LQC}} \star_{LQC} [f,H]_{\star_{LQC}}  \nonumber \\
    &=& \left\langle \frac{\partial f}{\partial t} \right\rangle + \frac{1}{i\hbar} \langle [f,H]_{\star_{LQC}} \rangle \,,
\label{eq:Ehren}
\end{eqnarray}
where the last line is a direct consequence of~(\ref{phasespacestar}).  Expression~(\ref{eq:Ehren}) corresponds to the Ehrenfest's theorem \cite{Zachos} within the context of LQC, and relates the time derivative of the expectation value of any observable defined on cylindrical functions in terms of star commutators, thus providing a correspondence principle from the deformation quantization perpective.

\section{Differential form of the star product}
\label{sec:diffstar}

It is also possible to determine the corresponding differential representation of the star product within the context of LQC. Let us first express the integral representation of the star product (\ref{star}) in terms of the Fourier transforms of the functions $f(c,\mu)$ and $g(c,\mu)$, we have
\begin{equation}\label{starFourier}
    (f\star_{LQC} g)(c,\mu)= \int_{\widehat{b\mathbb{R}}^2}  d\alpha \, d\beta \, \tilde{f}\left( \alpha, \mu + \frac{\hbar}{2}\beta \right) \tilde{g} \left( \beta, \mu - \frac{\hbar}{2}\alpha \right) h_{\alpha}(c) h_{\beta}(c) \,.
\end{equation}
In the standard case of Moyal's descritpion of Quantum Mechanics in phase space, one should perform a Taylor expansion of the functions appearing in the integral (\ref{starFourier}), making use of their smoothness and then, by using the Fourier convolution theorem one is able to write the star product as a noncommutative differential operator \cite{Zachos}. However, in the LQC context, since the second argument in the cylindrical functions $f(c,\mu)$ and $g(c,\mu)$ has a discrete nature, conduct a Taylor expansion is no longer valid. Instead, a discrete version of the Taylor formula may be 
adopted allowing to write the star product as a difference operator \cite{Zachos1}. By implementing the calculus of finite differences \cite{Ruzhansky}, the discrete Taylor expansion of a function $z\in \Cyl(b\mathbb{R}\times\widehat{b\mathbb{R}})$ reads 
\begin{equation}
\label{eq:discreteTaylor}
    z(c,\mu+\nu)=\sum_{k=0}^\infty \frac{1}{k!} \nu^{(k)}_{\mu_0} \triangle_{\mu,\mu_0}^k z(c,\mu)+r_{M}(\mu,\nu),
\end{equation}
where the operator $\triangle_{\mu,\mu_0}$ corresponds to the scaled forward difference operator defined by 
\begin{equation}
    \triangle_{\mu,\mu_0} z(c,\mu) = \frac{z(c,\mu+\mu_0)-z(c,\mu)}{\mu_0}, \qquad \triangle^k_{\mu,\mu_0} = \underbrace{\triangle_{\mu,\mu_0} \ldots \triangle_{\mu,\mu_0}}_{k \mbox{ times}} \,,
\end{equation}
while
\begin{equation}
    \nu^{(k)}_{\mu_0}= \nu (\nu-\mu_0) (\nu-2\mu_0) \ldots (\nu-(k-1)\mu_0) \,,
\end{equation}
and $r_{M}(\mu,\nu)$ denotes the remainder.
The parameter $\mu_{0}$, present in the discrete Taylor formula~(\ref{eq:discreteTaylor}), corresponds to a fixed length scale which restricts the dynamics to a lattice with regular spaced points, indicating that cylindrical functions on such space have support only on that lattice \cite{DQPolymer}. In view of the quantum nature of the geometry inherent to the loop quantization program, this parameter is related to a physical length whose value is associated with the minimum eigenvalue of the area operator in LQG and, in principle, it may be fixed by comological data \cite{APS}.

Then, by performing a discrete Taylor expansion in the second argument of the functions $\tilde{f}$ and $\tilde{g}$ in equation (\ref{starFourier}), we obtain
\begin{eqnarray}\label{stardiff}
  \mkern-50mu 
   (f\star_{LQC} g)(c,\mu) &=& \sum_{n=0}^{\infty} \sum_{k=0}^{n} \frac{1}{k!(n-k)!} \int_{\widehat{b\mathbb{R}}^2}  d\alpha \, d\beta \, \left(\frac{\hbar}{2}\beta \right)^{(n-k)}_{\mu_0} \left( -\frac{\hbar}{2} \alpha \right)^{(k)}_{\mu_0} \nonumber\\
    &&\times\triangle_{\mu,\mu_0}^{n-k} \tilde{f}(\alpha,\mu) \triangle_{\mu,\mu_0}^{k} \tilde{g}(\beta,\mu) h_{\alpha}(c) h_{\beta}(c) \\
    &=& \sum_{n=0}^{\infty} \sum_{k=0}^{n} \frac{1}{k!(n-k)!} \left( \frac{i\hbar}{2} \frac{d}{dc} \right)^{(k)}_{\mu_0} \triangle_{\mu,\mu_0}^{n-k} f(c,\mu) \left( -\frac{i\hbar}{2} \frac{d}{dc} \right)^{(n-k)}_{\mu_0} \triangle_{\mu,\mu_0}^k g(c,\mu), \nonumber
\end{eqnarray}
where the Fourier transform on $L^{2}(b\mathbb{R},dc)$, given by expression (\ref{FourierbR}), has been applied together with the identity
\begin{equation}
    \int_{\widehat{b\mathbb{R}}}  d\nu \, \nu^{(k)}_{\mu_0} \, \tilde{f}(\nu,\mu) h_\nu (c) = \left(-i\frac{d}{dc}\right)^{(k)}_{\mu_0} f(c,\mu).
\end{equation}
Equation (\ref{stardiff}) may be interpreted as the differential representation of the star product (\ref{star}) within the context of LQC and, as we can observe, it is given by a combination of a difference and a differential operator, in complete agreement with the properties of the space $\Cyl(b\mathbb{R}\times\widehat{b\mathbb{R}})$. It is worth mentioning that formula (\ref{stardiff}) was obtained in \cite{Stottmeister2} by employing almost-periodic pseudodifferential operators applied to $U(1)$ Khon-Nirenberg calculus with the aim to analyze adiabatic perturbations in LQG using the Born-Oppenheimer framework, while our approach is focused on the deformation quantization formalism. In particular, if we choose the functions $f(c,\mu)=h_{\mu}(c)$ and $g(c,\mu)=\mu$, after performing the series expansion, the representation of the star product (\ref{stardiff}) yields
\begin{equation}
    h_\mu (c) \star_{LQC} \mu = \mu h_\mu (c) + \frac{i\hbar}{2} \frac{d h_\mu (c)}{dc} \triangle_{\mu,\mu_0} \mu = \mu h_\mu (c) - \frac{\hbar}{2} \mu h_\mu (c),
\end{equation}
where $dh_\mu(c)/dc = i\mu h_\mu(c)$. Similarly, we can obtain
\beq
     \mu \star_{LQC} h_\mu(c) 
     & = &  \mu h_\mu (c) - \frac{i\hbar}{2} \frac{d h_\mu (c)}{dc} \triangle_{\mu,\mu_0} \mu = \mu h_\mu (c) + \frac{\hbar}{2} \mu h_\mu (c), \nonumber \\
     h_\mu(c) \star_{LQC} h_\nu(c) 
     & = &  h_\mu(c) h_\nu(c), \nonumber\\
     \mu \star_{LQC} \mu 
     & = &  \mu^2.
\end{eqnarray}
This means that the holonomy-flux algebra depicted in (\ref{commutator}) is fully recovered by means of the differential representation of the LQC star product. Moreover, it is straightforward to note that the first term in the series expansion (\ref{stardiff}) corresponds to the pointwise product between $f(c,\mu)$ and $g(c,\mu)$, as the rest of the terms in the series are multiplied by different powers of $\hbar$, which vanish in the limit $\hbar \to 0$, thereby the differential representation of the LQC star product determines a fiducial deformation of the usual pointwise product of cylindrical functions. .

\section{Conclusions}
\label{sec:conclu}

As it is well-known, the LQC program relies on a quantization on the Bohr compactification of the reals which naturally brings about discreteness, also incorporating the failure to simultaneously introduce well-defined position and momentum operators acting on the associated Hilbert space.  In consequence, we are confronted with a different representation that is clearly unequivalent to standard Schr\"odinger quantum mechanics.  Recently, a proposal to introduce the techniques of deformation quantization within the LQC formalism has been under analysis, being of particular relevance the manner in which the structures associated with the Bohr compactification adapt systematically to the non-commutative formulation emerging in the phase space quantization. Following previous results in this direction, as reported 
in~\cite{Sahlmann}, \cite{DQPolymer},~\cite{PolyW},~\cite{Quasi}, here we addressed the 
introduction of the integral representation of the star product of LQC.  
To that end, we started by considering, in close analogy to the 
formulation of the Moyal product, the Weyl quantization map explicitly defined 
for cylindrical functions defined on the Bohr compactification of the reals.  
As discussed in~\cite{Quasi}, such quantization map allows us to uniquely determine the holonomy and the momentum operators.
Therefore, we were able to identify the integral representation for this star product 
which, in particular, allowed us to demonstrate in an straightforward manner the non-commutative, bilinear and associative properties, 
which also, as described above, stands for a fiducial deformation of the 
usual pointwise product of cylindrical functions. 
Further, the introduced integral representation for the LQC star product granted us 
a direct comparison with the integral representation of the Moyal product, as the 
latter may be obtained by substituting the characters  of the Bohr compactification of the reals with the characters associated with the group of the real numbers.  
In this sense, our resulting integral representation settle a legitimate extension 
 of the Moyal product between functions defined on the Schwartz space by 
incorporating symbols defined on the space of cylindrical functions.

The integral representation for the LQC star product also favored the definition of a suitable star commutator that correctly reproduces the quantum representation of the holonomy-flux algebra for LQC \cite{report}.    Besides, by considering the appropriate classical limit, 
it was shown that this star commutator replicates the Poisson algebra emerging 
in the cosmological setup, thus introducing a one-parameter, associative deformation of the holonomy-flux classical algebra \cite{Bojowald}, \cite{report}.  In sight of this, we proposed a natural 
way to obtain the quantum dynamical evolution in LQC in terms of this star commutator
and a suitable Wigner quasi-probability distribution for LQC.
This dynamical equation extends the analogous in Moyal's approach by incorporating cylindrical functions within our formulation for LQC, and allowed us to reproduce in a simple manner the Ehrenfest theorem
within the LQC context, in agreement with previous literature \cite{Quasi}, \cite{Zachos}.

Further, by means of the discrete Taylor expansion we are also able to introduce an 
appropriate differential representation for the LQC star product, which is characterized by an 
appropriate combination of powers of differential and forward difference operators.   For this 
differential representation of the star-product, we have discussed the way in which it determines a fiducial deformation of the pointwise product for cylindrical functions, and we also have recovered in a transparent manner the holonomy-flux algebra of Loop Quantum Cosmology.

A relevant issue in our construction of both the integral and the differential representations
for the LQC star product is associated with the possibility to modify the standard machinery of deformation quantization in order to adapt it to the Bohr compactification of the reals.  This allowed us to recover
in an unambiguous manner certain results discussed previously in the literature where, starting from 
different mathematical principles,  a plethora of 
technical tools have to be implemented in order to obtain at the quantum level either the integral or the differential representations of a genuine operator product within the LQC framework \cite{Stottmeister1}, \cite{Stottmeister2}, \cite{Stottmeister3}.

Our claim is thus that the introduced integral and differential representations 
for the star product within the deformation quantization program for 
LQC may shed some light to a better comprehension of 
technical issues  ranging from the understanding of an appropriate semiclassical limit to 
the  evolution of quantum states, among 
others.  In particular, the former issue may be natural within the context of 
deformation quantization, while the latter may be appropriately tackled by 
considering the proposed LQC dynamical equation.

\section*{Acknowledgments}
The authors would like to acknowledge financial support from CONACYT-Mexico
under the project CB-2017-283838.\\

\noindent Data Availability Statement: No Data associated in the manuscript.


\bibliographystyle{unsrt}

\begin{thebibliography}{l}

\bibitem{Bojowald}Bojowald M., \emph{Loop Quantum Cosmology}, Living Rev.~Relativ.~{ \bf 8} 11 (2005), \texttt{arXiv:gr-qc/0601085}.

\bibitem{report}Ashtekar A. and Singh P., \emph{Loop Quantum Cosmology: A Status Report}, Class.~Quantum Grav.~{\bf 28} 213001 (2011), \texttt{arXiv:1108.0893 [gr-qc]}.

\bibitem{Rovelli}Rovelli C., \emph{Black hole entropy from loop quantum gravity}, Phys.~Rev.~Lett.~{\bf 77} 3288--91 (1996), \texttt{arXiv:gr-qc/9603063}.

\bibitem{Krasnov}Ashtekar A., Baez J., Corichi A. and Krasnov K., \emph{Quantum geometry and black hole entropy}, Phys.~Rev.~Lett.~{\bf 80} 904--7 (1998), \texttt{arXiv:gr-qc/9710007}.

\bibitem{Domangala}Domagala M. and Lewandowski J., \emph{Black-hole entropy from quantum geometry}, Class.~Quantum Grav.~{\bf 21} 5233--43 (2004), \texttt{arXiv:gr-qc/0407051}.

\bibitem{bounce}Bojowald M., \emph{Quantum nature of cosmological bounces}, Gen.~Relativ.~Gravit.~{\bf 40} 2659--83 (2008), \texttt{arXiv:0801.4001 [gr-qc]}.

\bibitem{Bojowald1}Bojowald M., \emph{Loop Quantum Cosmology and inhomogeneities}, Gen.~Rel.~Grav.~{\bf 38} 1771--1795 (2006), \texttt{arXiv:gr-qc/0609034}.

\bibitem{Agullo}Agullo I. and Singh P., \emph{Loop Quantum Cosmology: A Brief Review}, 100 Years of General Relativity (Loop Quantum Gravity: The First 30 Years vol 4) 
eds.~Ashtekar A. and Pullin J. (Singapore:
World Scientific, 2017), \texttt{arXiv:1612.01236 [gr-qc]}.

\bibitem{CRovelli}Rovelli C., \emph{Black Hole Evolution Traced Out with Loop Quantum Gravity}, Physics {\bf 11} 127 (2018), \texttt{arXiv:1901.04732 [gr-qc]}.

\bibitem{Perez}Amadei L., Liu H. and Perez A., \emph{Unitarity and Information in Quantum Gravity: A Simple Example}, Front. Astron. Space Sci. {\bf 8} 604047, \texttt{arXiv:1912.09750 [gr-qc]}.


\bibitem{Vandersloot}Singh P. and Vandersloot K., \emph{Semi-classical States, Effective Dynamics and Classical Emergence in Loop Quantum Cosmology}, Phys.~Rev.~{\bf D72} 084004 (2005), \texttt{arXiv:gr-qc/0507029}.

\bibitem{Flori1}Flori C. and Thiemann T., \emph{Semiclassical analysis of the Loop Quantum Gravity volume operator: I. Flux Coherent States}, (2008), \texttt{arXiv:0812.1537 [gr-qc]}.

\bibitem{ELivine}Livine E.~R., \emph{Some Remarks on the Semi-Classical Limit of Quantum Gravity}, Braz.~J.~Phys.~{\bf 35} 442 (2005), \texttt{arXiv:gr-qc/0501076}.

\bibitem{Dasgupta}Dasupta A., \emph{Semiclassical Loop Quantum Gravity and Black Hole Thermodynamics}, SIGMA {\bf 9} 013 (2013), \texttt{arXiv:1203.5119 [gr-qc]}.

\bibitem{Liu}Han M. and Liu H., \emph{Semiclassical limit of new path integral formulation from reduced phase space loop quantum gravity}, Phys.~Rev.~{\bf D102} 024083 (2020), \texttt{arXiv:2005.00988 [gr-qc]}.

\bibitem{MBojowald1}Bojowald M., \emph{Loop quantum gravity as an effective theory}, AIP Conf.~Proc.~{\bf 1483} 5 (2012), \texttt{arXiv:1208.1463 [gr-qc]}.

\bibitem{MBojowald2}Bojowald M. and Skirzewski A., \emph{Effective theory for the cosmological generation of structure}, Adv.~Sci.~Lett.~{\bf 1} 92 (2008), \texttt{arXiv:0808.0701 [astro-ph]}.

\bibitem{MBojowald3}Bojowald M., \emph{Effective Field Theory of Loop Quantum Cosmology}, Universe~{\bf 5} 44 (2019), \texttt{arXiv:1906.01501 [gr-qc]}. 

\bibitem{APerez}Perez A., \emph{On the regularization ambiguities in loop quantum gravity}, Phys.~Rev.~{\bf D73} 044007 (2006), \texttt{arXiv:gr-qc/0509118}.

\bibitem{Brahma}Brahma S., Ronco M., Amelino-Camelia G. and Marciano A., \emph{Linking loop quantum gravity quantization ambiguities with phenomenology}, Phys.~Rev.~{\bf D95} 044005, \texttt{arXiv:1610.07865 [gr-qc]}.

\bibitem{Navascues}Navascues B.~E. and Mena-Marugan G.~A., \emph{Quantization ambiguities and the robustness of effective descriptions of primordial perturbations in hybrid Loop Quantum Cosmology}, Class.~Quantum Grav.~{\bf 39} 015017, \texttt{arXiv:2102.00124 [gr-qc]}.




\bibitem{Critics}Bojowald M., \emph{Critical evaluation of common claims in Loop Quantum Cosmology}, Universe {\bf 6} 36 (2020), \texttt{arXiv:2002.05703 [gr-qc]}.

\bibitem{Sahlmann}Fewster C.~J. and Sahlmann H., \emph{Phase space quantization and Loop Quantum Cosmology: a Wigner function for the Bohr-compactified real line}, Class.~Quantum Grav.~{\bf 25} 225015 (2008), \texttt{arXiv:0804.2541 [math-ph]}.

\bibitem{Perlov}Perlov L., \emph{Uncertainty Principle in Loop Quantum Cosmology by Moyal Formalism}, J.~Math.~Phys.~{\bf 59} 032304 (2018), \texttt{arXiv:1610.06532 [gr-qc]}.

\bibitem{DQPolymer}Berra-Montiel J. and Molgado A., \emph{Polymer quantum mechanics as a deformation quantization}, Class.~Quantum Grav.~{\bf 36} 025001 (2019), \texttt{arXiv:1805.05943 [gr-qc]}.

\bibitem{PolyW}Berra-Montiel J., \emph{The Polymer representation for the scalar field: a Wigner functional approach}, Class.~Quantum Grav.~{\bf 37} 025006 (2020), \texttt{arXiv:1908.09194 [gr-qc]}.

\bibitem{Quasi}Berra-Montiel J. and Molgado A., \emph{Quasi-probability distributions in Loop Quantum Cosmology}, Class.~Quantum Grav.~{\bf 37} 215003 (2020), 
\texttt{arXiv:2007.01324 [gr-qc]}.

\bibitem{Tomography}Berra-Montiel J. and Molgado A., \emph{Tomography in Loop Quantum Cosmology}, Eur.~Phys.~J.~Plus {\bf 137} 283 (2022), \texttt{arXiv:2104.09721 [gr-qc]}. 

\bibitem{Stottmeister1}Stottmeister A. and Thiemann T., \emph{Coherent states, quantum gravity, and the Born-Oppenheimer approximation. I. General considerations}, J.~Math.~Phys.~{\bf 57} 063509 (2016), \texttt{arXiv:1504.02169 [math-ph]}.

\bibitem{Stottmeister2}Stottmeister A. and Thiemann T., \emph{Coherent states, quantum gravity and the Born-Oppenheimer approximation, II: Compact Lie Groups}, J.~Math.~Phys.~{\bf 57} 073501 (2016), \texttt{arXiv:1504.02170 [math-ph]}.

\bibitem{Stottmeister3}Stottmeister A. and Thiemann T., \emph{Coherent states, quantum gravity and the Born-Oppenheimer approximation, III: Applications to loop quantum gravity}, J.~Math.~Phys.~{\bf 57} 083509 (2016), \texttt{arXiv:1504.02171 [math-ph]}.

\bibitem{Bayen1}Bayen F., Flato M., Fronsda C., Lichnerowicz A. and Sternheimer D., \emph{Deformation theory and quantization I. Deformations of symplectic structures}, Ann.~Phys., NY {\bf 111} 61--110 (1978).

\bibitem{Bayen2}Bayen F., Flato M., Fronsdal C., Lichnerowicz A. and Sternheimer D., \emph{Deformation theory and quantization. II. Physical applications}, Ann.~Phys., NY {\bf 111} 111--51 (1978).

\bibitem{Kontsevich}Kontsevich M., \emph{Deformation quantization of poisson manifolds}, Lett.~Math.~Phys.~{\bf 66} 157--216 (2003), \texttt{arXiv:q-alg/9709040}.

\bibitem{Waldman}Waldmann S., \emph{Recent developments in deformation quantization}, in Proc. Regensburg
Conf.~2014 on Quantum Mathematical Physics, eds.~Finster F., Kleiner J., R\"oken C.
and Tolksdorf J. (Birkh\"auser, 2016) pp. 421--439, \texttt{arXiv:1502.00097 [math.QA]}.

\bibitem{Fredenhagen}Fredenhagen K. and Rejzner K., \emph{Perturbative construction of models of algebraic Quantum Field Theory}, in Advances in Algebraic Quantum Field Theory,
eds. Brunetti R., Dappiaggi C., Fredenhagen K. and Yngvason J. (Springer Cham,
Switzerland, 2015), pp. 31--74, \texttt{arXiv:1503.07814 [math-ph]}.

\bibitem{Compean}Garcia-Compean H., Plebansky J.~F., Przanowski M. and Turrubiates F.~J., \emph{Deformation quantization of classical fields}, Int.~J.~Mod.~Phys. {\bf A16} 2533--2558 (2001), \texttt{arXiv:hep-th/9909206}.

\bibitem{Compean2}Cordero R., Garcia-Compean H., and Turrubiates F.~J., \emph{Deformation quantization of
cosmological models}, Phys.~Rev. {\bf D83} 125030 (2011), \texttt{arXiv:1102.4379 [hep-th]}.

\bibitem{GA}Berra-Montiel J. and Molgado A, \emph{Deformation quantization of constrained systems: a group averaging approach}, Class.~Quantum Grav.~{\bf 37} 055009 (2020), \texttt{arXiv:1911.00945 [gr-qc]}.

\bibitem{coherent}Berra-Montiel J. and Molgado A., \emph{Coherent representation of fields and deformation quantization}, Int.~J.~Geom.~Meth.~Mod.~Phys.~{\bf 17} 11 2050166 (2020), \texttt{arXiv:2005.14333 [quant-ph]}.

\bibitem{path}Berra-Montiel J., \emph{Star product representation of coherent state path integrals}, Eur.~Phys.~J.~Plus~{\bf 906} (2020), \texttt{arXiv:2007.02483 [quant-ph]}.

\bibitem{tomographic}Berra-Montiel J. and Cartas R., \emph{Deformation quantization and the tomographic representation of quantum fields}, Int.~J.~Geom.~Meth.~Mod.~Phys. {\bf 17} 2050217 (2020), \texttt{arXiv:2005.14333 [quant-ph]}.

\bibitem{squeezed}Bianchi E., Guglielmon J., Hackl L. and Yokozimo N., \emph{Squeezed vacua in loop quantum gravity}, (2016), \texttt{arXiv:1605.05356 [gr-qc]}.

\bibitem{Corichi}Corichi A. and Montoya E., \emph{Coherent semiclassical states for loop quantum cosmology}, Phys.~Rev.~D~{\bf 84} 044021 (2011), \texttt{arXiv:1105.5081 [gr-qc]}.

\bibitem{Oriti}Oriti D., Pereira R. and Sindoni L., \emph{Coherent states in quantum gravity: a construction based on the flux representation of loop quantum gravity}, J.~Phys.~A:~Math.~Theor.~{\bf 45} 244004 (2012), \texttt{arXiv:1110.5885 [gr-qc]}.


\bibitem{Cosmas}Zachos C., \emph{Deformation Quantization: Quantum Mechanics lives and works in phase-space}, Int.~J.~Mod.~Phys.~A~{\bf 17} 297 (2002), \texttt{arXiv:hep-th/0110114}. 

\bibitem{Lee}Lee H.-W., \emph{Theory and application of the quantum phase-space distribution functions}, Phys.~Rep.~{\bf 259} 147--211 (1995).

\bibitem{Gradoni} Gradoni G., Creagh S.~C., Tanner G., Smartt C. and Thomas D.~W.~P., \emph{A phase-space approach for propagating field–field correlation
functions}, New J.~Phys.~{\bf 17} 093027 (2015), \texttt{	arXiv:1504.00507 [nlin.CD]}.




\bibitem{Folland}Folland G.~B., \emph{An Abstract Course in Harmonic Analysis}, 
2nd. edn. (London: Taylor and Francis, 2016).

\bibitem{Reiter}Reiter H. and Stegeman J.~D., \emph{Classical Harmonic Analysis and Locally Compact Groups} (London Mathematical Society Monographs) 2nd edn (Oxford: Clarendon, 2000).

\bibitem{Shubin}Shubin~M.~A., \emph{Almost periodic functions and partial differential operators}, Russian~Math.~Surveys~{\bf 33} 2 (1978). 

\bibitem{Deitmar}Deitmar A. and Echterhoff S., \emph{Principles of Harmonic Analysis}, 2nd. edn. (Switzerland: Springer, 2014).

\bibitem{Chojnacki}Chojnaki W., \emph{Almost periodic Schr\"odinger operators in $L^{2}(b\mathbb{R})$ whose point spectrum is not all of the spectrum}, J.~Fun.~Anal.~{\bf 65} 236 (1986). 

\bibitem{Ashtekar}Ashtekar A., Bojowald M. and Lewandowski J, \emph{Mathematical structure of Loop Quantum
cosmology}, Adv.~Theor.~Math.~Phys.~{\bf 7} 233 (2003), \texttt{arXiv:gr-qc/0304074}.

\bibitem{Singh}Ashtekar A., Corichi A. and Singh P., \emph{Robustness of key features of Loop Quantum Cosmology}, Phys.~Rev. {\bf D77} 024046 (2008), \texttt{arXiv:0710.3565 [gr-qc]}.

\bibitem{Takhtajan}Takhtajan L.~A., \emph{Quantum Mechanics for Mathematicians (Graduate Studies in Mathematics, Vol 95)} (Rhode Island: American Mathematical Society, 2008).


\bibitem{Zachos}Zachos C.~K., Fairlie D.~B. and Curtright T.~L., \emph{Quantum Mechanics in Phase Space: An Overview with Selected Papers}, (Singapore: World Scientific, 2005).


\bibitem{Hossain}Hossain G.~M., Husain V. and Seahra S.~S., \emph{Background independent quantization and the uncertainty principle}, Class.~Quantum~Grav.~{\bf 27} 165013 (2010), \texttt{arXiv:1003.2207 [gr-qc]}.

\bibitem{Zachos1}Curtright T.~L., Fairlie D.~B. and Zachos C.~K., \emph{Features of Time-independent Wigner Functions}, Phys.~Rev.~{\bf D58} 025002 (1998), \texttt{arXiv:hep-th/9711183}.

\bibitem{Ruzhansky}Ruzhansky M. and Turunen V., \emph{Pseudo-Differential Operators and Symmetries: Background Analysis and Advanced Topics}, (Basel: Birkh\"auser Verlag, 2010)

\bibitem{APS}Ashtekar A., Pawlowski P. and Singh P., \emph{Quantum Nature of the Big Bang: Improved dynamics}, Phys.~Rev.~{\bf D74} 084003 (2006), \texttt{arXiv:gr-qc/0607039}.



\end{thebibliography}

\end{document}